\documentclass{ws-procs9x6}
\def\IR{\relax{\rm I\kern-.18em R}}
\def\o#1{\overline{#1}}

\newcommand{\ba}{\begin{eqnarray}}
\newcommand{\ea}{\end{eqnarray}}

\begin{document}

\title{String unification of gauge couplings with intersecting D-branes}

\author{Ralph Blumenhagen}

\address{         DAMTP, Centre for Mathematical Sciences \\
                  Wilberforce Road\\ 
                  Cambridge CB3 0WA, United Kingdom\\
                  e-mail:R.Blumenhagen@damtp.cam.ac.uk }
                  



\maketitle

\abstracts{After reviewing the general structure of supersymmetric
intersecting brane world models, we discuss the 
issue of stringy gauge coupling unification for a natural
class of MSSM-like models.  }

\section{Introduction}

During the last years string compactifications with intersecting 
D-branes \cite{Blumenhagen:2000wh,Angelantonj:2000hi,Aldazabal:2000cn,Aldazabal:2000dg,Blumenhagen:2000ea,Ibanez:2001nd,Blumenhagen:2001te,Cvetic:2001nr,Kokorelis:2002ip,Blumenhagen:2002wn,Ellis:2002ci}
 have been of particular interest to the 
string phenomenology community  for their
appealing  Standard Model-like  features such as chiral
fermions \cite{Berkooz:1996km,Ohta:1997fr}, 
family replication and Standard Model gauge groups.

From the model building point of view, having
a low string scale scenario in mind,  most effort went
into the construction of non-supersymmetric string models,
which, as many non-supersymmetric string models do, suffer
from perturbative instabilities like tachyons and/or
uncancelled disc tadpoles for some of the scalar  fields including the dilaton 
\cite{Dudas:2000ff,Blumenhagen:2000dc,Blumenhagen:2001te}.

Clearly, from the stringy point of view it is much more
convincing  to start with a supersymmetric intersecting brane 
configuration, where the string scale is unconstrained
and can be everywhere between the weak and the Planck scale. 
This set-up is closer to the heterotic string, which 
was the string phenomenologists favorite branch of the M-theory moduli space
before the concept of D-branes was  introduced. 
Unfortunately, as concrete model building attempts have
revealed, \cite{Cvetic:2001nr,Blumenhagen:2002gw,Larosa:2003mz,Honecker:2003vq,Cvetic:2003xs}
realistic supersymmetric intersecting 
D-brane models are much more difficult to uncover,
as supersymmetry appears to be a fairly rigid constraint.

Despite this lack of simple concrete realizations, 
one might think about general aspects of this class
of models. For instance, the discussion of the general
structure of three and four point-functions 
\cite{Cremades:2003qj,Cvetic:2003ch,Abel:2003fk,Abel:2003vv}
and the discussion of the proton lifetime in intersecting brane world
models (IBWs) \cite{Klebanov:2003my} are nice
examples of this general approach. 

In this letter we review another recent attempt, which deals with  
the issue of gauge coupling unification in a very natural
class of MSSM-like intersecting brane world models 
\cite{Blumenhagen:2003jy}.
After reviewing briefly the general structure of IBWs and introducing
a natural  class of MSSM like models,
we discuss how perturbative gauge coupling unification might arise.
Note, that in this article we are using the term ``unification''
in the loose sense, that the values of the gauge couplings at
the string scale are consistent with the stringy tree level predictions.
In IBWs this does not necessarily mean, that the couplings really meat
at some scale, as in the usual field theoretic GUT scenario.

\section{The general set-up}

We consider Type IIA string theory compactified on an 
orientifold background of the form \cite{Blumenhagen:2002wn}
\begin{equation}
{X} = \IR^{3,1} \times {{M} \over \Omega \o\sigma},
\end{equation}
where  ${M}$ is a Calabi-Yau manifold and
$\o\sigma$ denotes an anti-holomorphic involution.

The fixed point  locus of $\o\sigma$ gives rise to  orientifold
O6 planes, whose R-R charge must be canceled by introducing D6-branes
in the background. The easiest possibility, which has been mostly studied
in the literature, is to place the D6-branes directly on top of or at least
parallel to the orientifold planes. 
The whole philosophy about intersecting brane worlds is to give up
this restriction and allow the D6 branes to occupy  more general
 positions. 
This has the effect that they will  intersect each other non-trivially leading
to both chirality \cite{Berkooz:1996km} 
and in general supersymmetry breaking. 

The massless modes are localized on defects in the ten-dimensional
space-time of different dimensionality.
Since the internal space is compact, we have to cancel 
the R-R tadpoles. 
In local coordinates the anti-holomorphic involution can be written as
$\o\sigma:z_i\to \o{z}_i$ and the 
fixed locus  is a special Lagrangian (sLag) 3-cycle.
Now we introduce  general D6-branes wrapped on  homology cycles
$\pi_a$ and their $\Omega\o\sigma$ images $\pi'_a$.
For such A-type D-branes wrapping sLag cycles with vanishing 
gauge field strength ${ F}=0$ the 
R-R 7-form  charge on the compact manifold ${ M}$ vanishes
if the total homology class vanishes
\begin{equation}
 \sum_a  N_a\, (\pi_a + \pi'_a)
          -4\,   \pi_{O6}=0.
\end{equation}
In a supersymmetric configuration all branes are calibrated with 
respect to the same 3-form as the orientifold plane implying
that the scalar potential vanishes as the consequence of 
the R-R tadpole condition.
Note, that the scalar potential only depends on the complex structure 
of ${ M}$.

As we show in Table \ref{tab1} the chiral massless spectrum is 
entirely determined by the topological intersection numbers between pairs
of D6-branes. 

\begin{table}[th]
\tbl{Massless 4D chiral fermion spectrum.\vspace*{1pt}}
{\footnotesize
\begin{tabular}{|c|r|r|}
\hline
{} &{} &{}\\[-1.5ex]
Sector & Rep. & Number \\[1ex]
\hline
{} &{} &{} \\[-1.5ex]
$a\,a'$ & $A_a$ &
${1\over 2}\left(\pi'_a\circ\pi_a+\pi_{O6}\circ\pi_a\right)$ \\[1ex]
$a\, a'$ & $S_a$ & ${1\over
  2}\left(\pi'_a\circ\pi_a-\pi_{O6}\circ\pi_a\right)$ \\[1ex]
$a\,b$ & $(\o{N}_a,N_b)$ & $\pi_{a}\circ \pi_b$  \\[1ex]
$a\, b'$ & $({N}_a,{N}_b)$  & $\pi'_{a}\circ \pi_b$ \\[1ex]
\hline
\end{tabular}\label{tab1} }
\end{table}

The non-abelian gauge anomalies cancel automatically and mixed
$U(1)_a-SU(N)^2_b$ anomalies are canceled by a  generalized Green-Schwarz
mechanism  involving dimensionally reduced R-R forms.

\section{Gauge couplings}

In contrast to the heterotic string, here
each gauge factor comes with its own gauge coupling,
which at string tree-level can be deduced from the
Dirac-Born-Infeld action to be 
\begin{equation}
 {4\pi \over g_a^2}={M_s^3\, V_{a} \over (2\pi)^3\, g_{st}\, 
\kappa_a}
\end{equation}
with $\kappa_a=1$ for $U(N_a)$ and $\kappa_a=2$ for $SP(2N_a)/SO(2N_a)$.

By dimensionally reducing the type IIA gravitational action one can  
similarly express the Planck mass in terms of stringy parameters
($M_{pl}=(G_N)^{-{1\over 2}}$)
\begin{equation}
{M_{pl}^2}={8\, M_s^8\, V_{6} \over (2\pi)^6\, g^2_{st}}.
\end{equation}

A very natural way to embed the Standard Model into intersecting
brane worlds \cite{Cremades:2002va} is to start 
with four stacks of D6-branes giving 
rise to the initial gauge symmetry
$U(3)\times SP(2) \times U(1)\times U(1)$ (with $SP(2)\simeq SU(2)$).
Requiring that the chiral spectrum of this model agrees with 
the three generation Standard Model augmented by a right
handed neutrino uniquely fixes the 
intersection numbers between pairs of D-branes. 
The hypercharge is given by the following linear combination
of $U(1)$ factors
\begin{equation}
Q_Y={1\over 3}Q_a-Q_c-Q_d .
 \nonumber
\end{equation}
leading to the gauge coupling
\begin{equation}
{1\over \alpha_Y}={1\over 6}{1\over \alpha_a} +
                                  {1\over 2}{1\over \alpha_c} +
                                  {1\over 2}{1\over \alpha_d} .
\end{equation}
From the stringy point of view a natural subclass of such models
show  some more symmetry among  the four appearing gauge couplings.
Since the intersection numbers of the first and fourth stack of
branes are identical and both branes are calibrated, in the simplest 
case on gets that the internal volumes agree too, $V_a=V_d$.
Employing the condition that the Green-Schwarz terms still allow
for a massless hypercharge, allows one to deduce
that the third stack satisfies $\pi_c'=\pi_c$. 
Therefore, at the bottom of this simple realization
there lies an extended Pati-Salam like model 
\begin{equation}
U(4)\times SU(2)\times SU(2).
\end{equation}
where in the following we will also assume that the 
two gauge couplings of the two $SU(2)$ factors
are identical (which in the simple toroidal models
discussed in \cite{Cremades:2002va,Kokorelis:2003jr} 
is a consequence of supersymmetry).
This allows us to derive the following string tree level
Pati-Salam like relation among the gauge couplings
\begin{equation}
{1\over \alpha_Y}={2\over 3}{1\over \alpha_{s}} +
                                  {1\over \alpha_{w}}.
\end{equation}

\section{One loop corrections to the gauge couplings}

Assuming that all gauge couplings are in the perturbative
regime, we now study the one-loop running
of the three gauge couplings and determine whether
for appropriately chosen
string scale the string tree level relation can 
be consistent with the known values of the Standard 
Model gauge  couplings at the weak scale $M_w$.
The tree level relation at the string scale yields 
\begin{equation}
   {2\over 3}{1\over \alpha_{s}(M_w)}+{2\sin^2\theta_w(M_w)-
                1 \over \alpha(M_w)}={B\over 2\pi}\, 
     \ln\left({M_w\over M_s}\right) 
\end{equation}
with
\begin{equation}
  B={2\over 3}\,b_s+ b_w-b_y.
\end{equation}
where $b_s$, $b_w$ and $b_y$ are the beta function coefficient
for the strong, weak and hypercharge gauge coupling  respectively. 
Note, that the resulting value of the unification scale
only depends on the combination $B$ of the beta-function coefficients.

In general besides the chiral matter  string theory contains also 
additional vector-like matter. 
This is also localized on higher dimensional intersection
loci of the $D6$ branes and also comes with multiplicities $n_{ij}$ with
$i,j\in\{a,b,c,d\}$. 
Assuming the most general vector-like matter in bifundamental, 
(anti-)symmetric and adjoint representations of the gauge group, 
one finds the following contribution to $B$
\ba
  B&=&{12}  - 2\, n_{aa} - {4}\, n_{ab} + 2\, n_{a'c} 
         + 2\, n_{a'd} - 2\, n_{bb}  +
             2\, n_{c'c} \nonumber \\
  &&  +  2\, n_{c'd}  + 2\, n_{d'd}. 
 \ea
$B$ does not depend on the number of weak Higgs multiplets $n_{bc}$.
Depending on $B$, one finds the discretuum  of  
 unification scales displayed in Table 2. 
\begin{table}[th]
\tbl{String scales.\vspace*{1pt}}
{\footnotesize
\begin{tabular}{|c|r|}
\hline
{} &{}\\[-1.5ex]
{\bf B} & ${\bf M_s}$[GeV]  \\[1ex]
\hline
{} &{} \\[-1.5ex]
$18$ & $3.36\cdot 10^{11}$ \\[1ex]
$16$ & $5.28\cdot 10^{12}$ \\[1ex]
$14$ & ${ 1.82\cdot 10^{14}}$ \\[1ex]
$12$ & $2.04\cdot 10^{16}$ \\[1ex]
$10$ & $1.51\cdot 10^{19}$ \\[1ex]
$8$ &  $3.06\cdot 10^{23}$ \\[1ex]
\hline
\end{tabular}\label{tab2} }
\end{table}

\section{Examples}

For the MSSM one has the well known values  
  $(b_s,b_w,b_y)=(3,-1,-11)$, i.e $B=12$ and
the unification scale is the usual GUT scale
\begin{equation}
M_X=2.04\cdot 10^{16}\, {\rm GeV}.
\end{equation}
Assuming $g_{st}=g_X$, for the internal sizes of the entire
Calabi-Yau manifold and the two 3-cycles where the
$SU(3)$ and $SU(2)$ branes are wrapped on, one obtains
\begin{equation}
M_s R=5.32, \quad M_s R_s=2.6, \quad M_s R_w=3.3.
\end{equation}
The running is shown in figure 1.
Note, that in contrast to the heterotic string, where one had
to invoke one-loop heavy threshold corrections at the string
scale, here 
unification of gauge couplings can be achieved with just the
tree level gauge couplings and appropriately chosen internal
radii. 

\begin{figure}[th]
\centerline{\epsfxsize=6cm\epsfbox{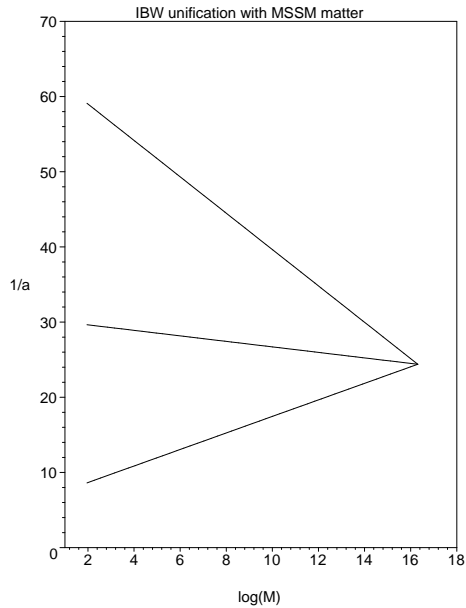}}   
\caption{Running couplings for MSSM matter (B=12). \label{inter}}
\end{figure}

As a second example, we consider  a model with a second weak Higgs
field, i.e.  $n_{bc}=1$,   so that we still
get $B=12$ but with $(b_s,b_w,b_y)=(3,-2,-12)$.
As shown in the left picture of figure 2, 
the gauge couplings 
still "unify" at  the usual GUT scale, but
do not really meet there.


Finally, let us discuss the case with  $B=10$, where one gets
the intriguing value
\begin{equation}
      {M_{s}\over M_{pl}}=1.24\sim \sqrt{\pi\over 2}. \nonumber
\end{equation}
Choosing for instance vector-like adjoint matter $n_{aa}=1$,
the beta-function coefficients read 
$(b_s,b_w,b_y)=(0,-1,-11)$
leading to the running shown in the right picture in figure 2. 
The couplings at the string  scale turn out to be
\begin{equation}
  \alpha_s(M_s)=0.117, \quad \alpha_w(M_s)=0.043, \quad
                  \alpha_Y(M_s)=0.035. \nonumber
\end{equation}
and the Weinberg angle at the string scale reads 
$\sin^2 \theta_w(M_s)=0.445$.
For the scales  of the overall Calabi-Yau volume and the 3-cycles
we obtain
\begin{equation}
M_s R=0.6, \quad M_s R_s=1.9,\quad M_s R_w=3.3 . \nonumber
\end{equation}

\begin{figure}[th]
\centerline{\hbox{\epsfxsize=6cm\epsfbox{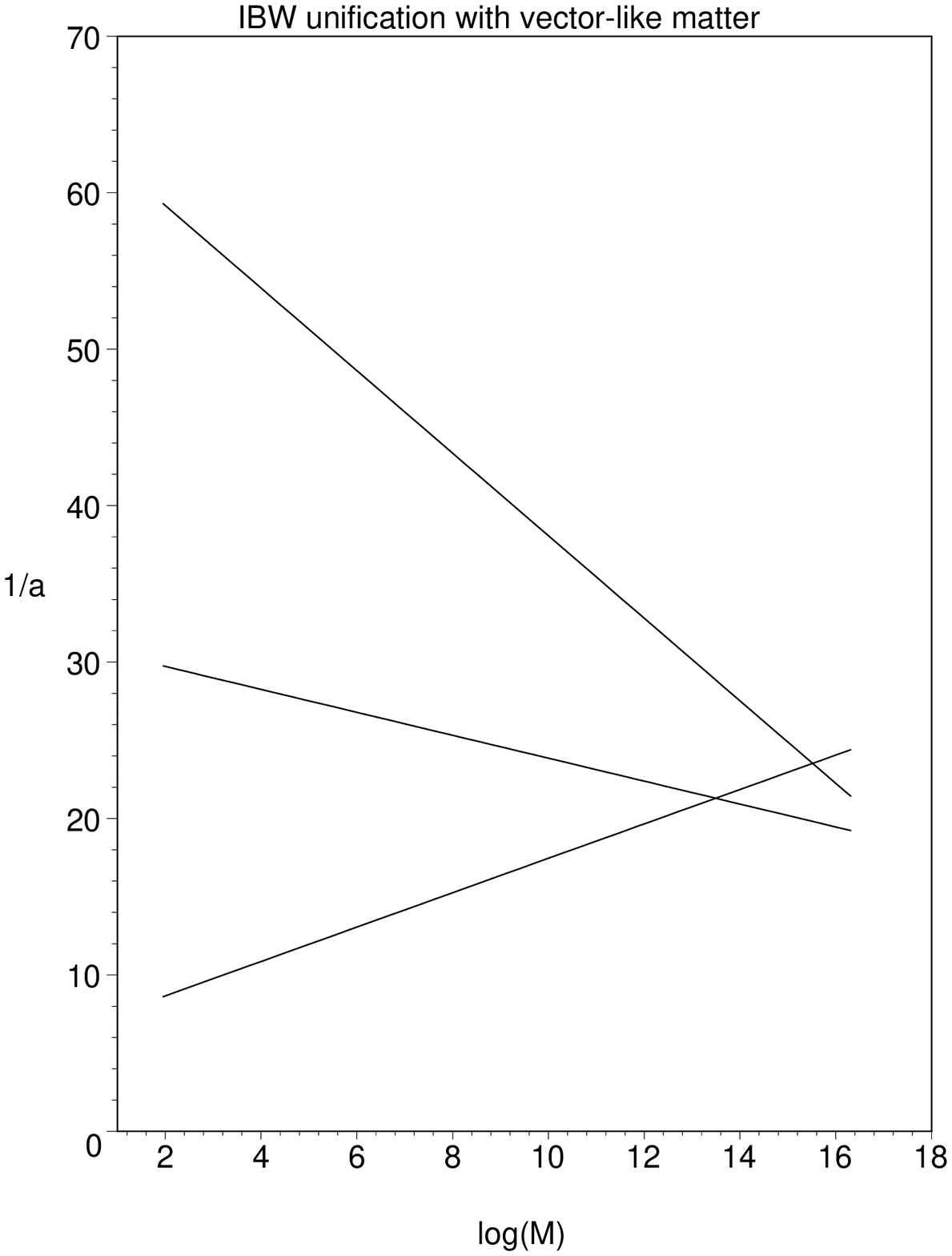}
             \epsfxsize=6cm\epsfbox{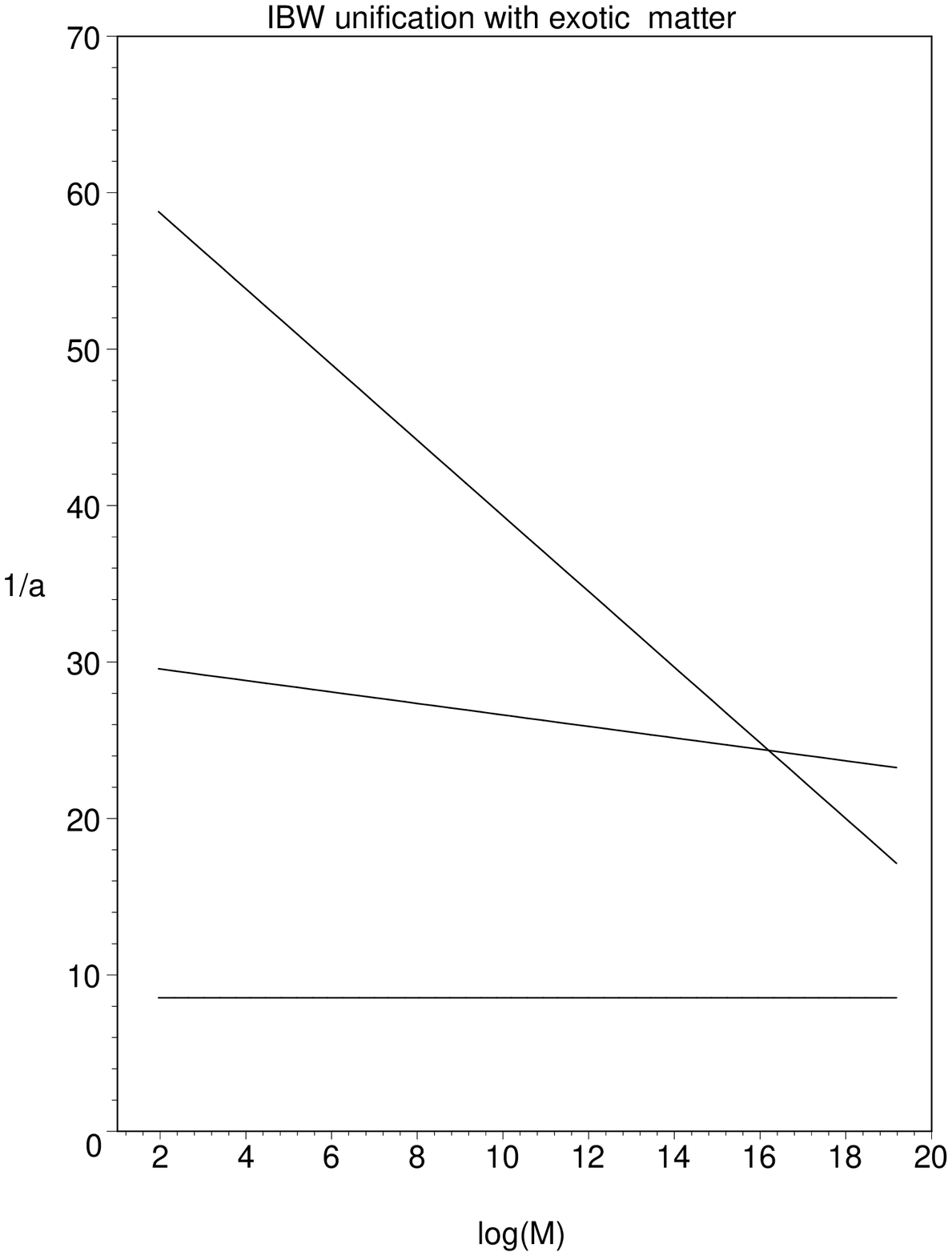} } }   
\caption{Gauge couplings for models with exotic matter, 
left: (B=12), right:   (B=10). \label{interb}}
\end{figure}

\section{Outlook}

We have shown that a general class of realistic IBWs 
feature perturbative gauge coupling unification. 
The burning problem remains to really construct
concrete intersecting brane models, which fall into
the class of models discussed in this letter.
For a concrete model, it would be interesting to analyze
what the effect of the one-loop threshold corrections 
\cite{Lust:2003ky} would be.

\section*{Acknowledgments}
I would like to thank D. L\"ust and S. Stieberger  for 
collaboration on the work presented at the "String Phenomenology 2003"
conference, Durham, 29 July - 4 August 2003. 
R.B. is supported by a PPARC Advanced Fellowship.

\end{document}